\documentclass[12pt,thmsa]{article}
\usepackage{sw20aip}

\input tcilatex
\begin{document}

\title{Multi-soliton Solution of the Integrable Coupled Nonlinear Schr\"{o}dinger
Equation of Manakov Type}
\author{Freddy P. Zen$^{1)\thanks{%
email: fpzen@bdg.centrin.net.id}}\,\,$and \thinspace Hendry I. Elim$^{1),2)%
\thanks{%
email: hendry202@cyberlib.itb.ac.id}}$ \and 1) Theoretical High Energy
Physics Group, \and \thinspace \thinspace \thinspace \thinspace \thinspace
\thinspace Theoretical Physics Lab.,\thinspace Department of
Physics,\thinspace \and \thinspace \thinspace \thinspace \thinspace
\thinspace \thinspace \thinspace Institute of Technology Bandung,\thinspace
Bandung,\thinspace Indonesia \and 2) Department of Physics,\thinspace
Pattimura University, \and \thinspace \thinspace \thinspace \thinspace
\thinspace \thinspace \thinspace Ambon,\thinspace \thinspace Indonesia}
\maketitle

\begin{abstract}
The general multi-soliton solution of the integrable coupled nonlinear
Schr\"{o}dinger equation (\textbf{NLS}) of Manakov type\thinspace
is\thinspace investigated by using Zakha- rov-Shabat (\textbf{ZS}) scheme.
We get the bright and dark multi-soliton solution using inverse scattering
method of \textbf{ZS} scheme. Elastic and inelastic collision of $N$%
-solitons solution of the equation are also discussed. 
\[
\]
PACS number(s): 42.65Sf
\end{abstract}

\section{Introduction}

The integrable coupled nonlinear Schr\"{o}dinger equation of Manakov type is
widely used in recent developments in the field of optical solitons in
fibers. The use and applications of the equation is to explain how the
solitons waves transmit in optical fiber, what happens when the
interaction\thinspace among optical solitons influences directly the
capacity and quality of communication and so\thinspace on.$^{1-4}$

Some of exact solutions have been derived for the system related to the
Manakov type equation with various methods.$^{5-10}$ Here, we will
investigate the general solution of the following equation which is closely
related to the Manakov type equation: 
\begin{eqnarray}
iq_{1_x}+\frac{\left( \alpha +\delta \right) }{\alpha \left( \alpha -\delta
\right) }q_{1_{tt}}+2\mu \left( \left| q_1\right| ^2+\left| q_2\right|
^2\right) q_1 &=&0  \nonumber \\[0.01in]
iq_{2_x}+\frac{\left( \alpha +\delta \right) }{\alpha \left( \alpha -\delta
\right) }q_{2_{tt}}+2\mu \left( \left| q_1\right| ^2+\left| q_2\right|
^2\right) q_2 &=&0,  \tag{1.1}
\end{eqnarray}
where $q_1$ and $q_2$ are slowly varying envelopes of the two interacting
optical modes, the variables $x$ and $t$ are the normalized distance and
time, and $\alpha $ and $\delta $ are arbitrary real values which obey $%
\alpha \neq \delta $ and $\alpha \neq -\delta $. On the other hand,
parameter $\mu $ is defined as follows 
\begin{equation}
\mu =\pm \frac{\alpha ^2-\delta ^2}{\alpha ^2\delta },  \tag{1.2}
\end{equation}
which is confirmed in section II.

Recently, in Ref.$11$, Radhakrishnan and Lakshmanan have derived the bright
and dark multi-soliton solution of the related system in eq.$(1.1)$ using
Hirota method. However in this paper, we will investigate and provide
another method to solve the related problem using Zakharov-Shabat inverse
scattering method. We show that it is possible when the Zakharov-Shabat (%
\textbf{ZS}) scheme (in its final form) is expressed solely in terms of
certain operators. Based on the choice of the operator, we find that there
is a multi-soliton solution of eq.$(1.1)$. We also find that there is an
elastic and inelastic collision of the bright and dark multi-soliton.

This paper is organized as follows. In section II, we will perform the 
\textbf{ZS} scheme for eq.$(1.1)$ and the arbitrary real parameter $\mu \,$%
(eq.$(1.2)$). In section III, we will solve the bright and dark
multi-soliton solution. We will compare the bright one and two soliton
solution with the results in Ref.$8$ and Ref.$10$. Section IV is devoted for
discussions and conclusions. In this section, we also state that there must
be an integrable$\,$vector\textbf{\ NLS} system.

\section{ZS \thinspace Scheme\thinspace for\thinspace the
Integrable\thinspace \thinspace \thinspace Coupled \thinspace \textbf{NLS}
Equation of Manakov Type}

We start by choosing the following two operators related to Zakharov-Shabat (%
\textbf{ZS}) scheme 
\begin{equation}
\Delta _0^{(1)}\,=\,I\left( i\alpha \frac \partial {\partial x}\,-\,\frac{%
\partial ^2}{\partial t^2}\right) ,  \tag{2.1a}
\end{equation}
$\,$and 
\begin{equation}
\Delta _0^{(2)}\,=\,\left( 
\begin{tabular}{lll}
$\alpha $ & $0$ & $0$ \\ 
$0$ & $\beta $ & $0$ \\ 
$0$ & $0$ & $\gamma $%
\end{tabular}
\right) \frac \partial {\partial t},  \tag{2.1b}
\end{equation}
where $\alpha $,$\,\beta \,$and $\gamma \,$ are arbitrary real values, and $%
I $ is the 3x3 unit matrix. We can then define the following\thinspace
operators by using this scheme related to inverse scattering techniques$%
^{12-14}$%
\begin{equation}
\Delta ^{(1)}\,=\,\,\Delta _0^{(1)}+\,U\left( t,x\right) ,  \tag{2.2a}
\end{equation}
and 
\begin{equation}
\Delta ^{(2)}\,=\,\Delta _0^{(2)}+\,V\left( t,x\right) .  \tag{2.2b}
\end{equation}
Here operators $\Delta ^{(i)}$,($i\,$=\thinspace 1,\thinspace 2) satisfy the
following equation 
\begin{equation}
\Delta ^{(i)}\left( I\,+\,\Phi _{+}\right) \,=\,\left( I\,+\,\Phi
_{+}\right) \Delta _0^{(i)},  \tag{2.3}
\end{equation}
where the Volterra integral operators $\Phi _{\pm }\left( \psi \right) \,$%
\thinspace are defined according to equation 
\begin{equation}
\Phi _{\pm }\left( \psi \right) \,=\,\int\limits_{-\infty }^\infty k_{\pm
}\left( t,z\right) \psi \left( z\right) dz.  \tag{2.4}
\end{equation}

We now\thinspace suppose that operators\thinspace $\Phi _F\left( \psi
\right) \,$and\thinspace \thinspace $\Phi _{\pm }\left( \psi \right) \,$ are
related to the following operator\thinspace identity 
\begin{equation}
\left( I\,+\,\Phi _{+}\right) \,\left( I\,+\,\Phi _F\right) \,=\left(
I\,+\,\Phi _{-}\right) ,  \tag{2.5}
\end{equation}
where\thinspace the integral operator $\Phi _F\left( \psi \right) \,$%
\thinspace is 
\begin{equation}
\Phi _F\left( \psi \right) \,=\,\int\limits_{-\infty }^\infty F\left(
t,z\right) \psi \left( z\right) dz.  \tag{2.6}
\end{equation}
Both $k_{+}\,$and$\,\,F\,\,$in eq.(2.4) and (2.6) are the 3x3 matrices
chosen as follows 
\begin{equation}
k_{+}=\,\left( 
\begin{tabular}{lll}
$a\left( t,z;x\right) $ & $q_1\left( t,z;x\right) $ & $q_2\left(
t,z;x\right) $ \\ 
$\pm q_1^{*}\left( t,z;x\right) $ & $d\left( t,z;x\right) $ & $e\left(
t,z;x\right) $ \\ 
$\pm q_2^{*}\left( t,z;x\right) $ & $f\left( t,z;x\right) $ & $g\left(
t,z;x\right) $%
\end{tabular}
\right) ,  \tag{2.7a}
\end{equation}
and 
\begin{equation}
F\,=\,\left( 
\begin{tabular}{lll}
$\,\,\,\,\,\,\,\,\,\,\,\,0$ & $A_n\left( t,z;x\right) $ & $B_n\left(
t,z;x\right) $ \\ 
$A_n^{*}\left( t,z;x\right) $ & $\,\,\,\,\,\,\,\,\,\,\,\,0$ & $%
\,\,\,\,\,\,\,\,\,\,\,\,\,0$ \\ 
$B_n^{*}\left( t,z;x\right) $ & $\,\,\,\,\,\,\,\,\,\,\,\,0$ & $%
\,\,\,\,\,\,\,\,\,\,\,\,\,0$%
\end{tabular}
\right) .  \tag{2.7b}
\end{equation}
Here $a$, $q_1$, $q_2$, $\pm q_1^{*}$, $\pm q_2^{*}$, $d$, $e$, $f$, $g$, $%
A_n$, $A_n^{*}$, $B_n$ and $B_n^{*}$ are parameters which will be calculated
in section III.

In eq.$(2.5)$, we have assumed that $\left( I\,+\,\Phi _{+}\right) \,\,$is
invertible, then 
\begin{equation}
\,\left( I\,+\,\Phi _F\right) \,=\left( I\,+\,\Phi _{+}\right) ^{-1}\left(
I\,+\,\Phi _{-}\right) ,  \tag{2.8}
\end{equation}
so that operator $\left( I\,+\,\Phi _F\right) \,$ is factorisable. From eq.$%
(2.5)$, we derive Marchenko matrix equations, 
\begin{equation}
k_{+}\left( t,z;x\right) +\,F\left( t,z;x\right) \,+\,\int\limits_t^\infty
k_{+}\left( t,t^{\prime };x\right) F\left( t^{\prime },z;x\right) dt^{\prime
}\,=\,0,\,\,\,\,\text{\thinspace \thinspace for}\,\,z\,>\,t\,\,,  \tag{2.9a}
\end{equation}
and 
\begin{equation}
k_{-}\left( t,z;x\right) \,=\,F\left( t,z;x\right) \,+\,\int\limits_t^\infty
k_{+}\left( t,t^{\prime };x\right) F\left( t^{\prime },z;x\right) dt^{\prime
}\,,\,\,\,\,\,\,\text{for}\,\,z\,<\,t.\,  \tag{2.9b}
\end{equation}
In eq.$(2.9b)$,\thinspace $k_{-}\,\,$is obviously defined in terms of $%
k_{+}\,$and\thinspace $F\,$.\thinspace Both eq.$(2.9a)\,$and $(2.9b)$%
\thinspace require $F$, and $F$ is supplied by the solution of equations : 
\begin{equation}
\left[ \Delta _0^{(1)},\Phi _F\right] =0,  \tag{2.10a}
\end{equation}
and 
\begin{equation}
\left[ \Delta _0^{(2)},\Phi _F\right] =0.  \tag{2.10b}
\end{equation}
After a little algebraic manipulation, we get 
\begin{equation}
\,i\alpha F_x\,+\,F_{zz}\,-\,F_{tt}\,=\,0,  \tag{2.11a}
\end{equation}
and 
\begin{equation}
\,\,\,\,\left( 
\begin{tabular}{lll}
$\alpha $ & $0$ & $0$ \\ 
$0$ & $\beta $ & $0$ \\ 
$0$ & $0$ & $\gamma $%
\end{tabular}
\right) F_t\,+\,F_z\left( 
\begin{tabular}{lll}
$\alpha $ & $0$ & $0$ \\ 
$0$ & $\beta $ & $0$ \\ 
$0$ & $0$ & $\gamma $%
\end{tabular}
\right) \,=\,0,  \tag{2.11b}
\end{equation}
where $F_t\equiv \frac{\partial F}{\partial t}$, $F_z\equiv \frac{\partial F%
}{\partial z}$, etc.

$U\left( t,x\right) \,$and $V\left( t,x\right) \,$can be found by solving eq.%
$(2.3)\,$in which we have substituted eq.($2.7a$) and ($2.4$) (for $k_{+})\,$%
to that equation: 
\begin{equation}
V\left( t,x\right) \,=\,\left( 
\begin{tabular}{lll}
$\,\,\,\,\,\,\,\,\,\,\,\,\,\,\,\,\,0$ & $\left( \alpha -\beta \right) q_1$ & 
$\left( \alpha -\gamma \right) q_2$ \\ 
$\pm \left( \beta -\alpha \right) q_1^{*}$ & $\,\,\,\,\,\,\,\,\,\,\,\,0$ & $%
\left( \beta -\gamma \right) e$ \\ 
$\pm \left( \gamma -\alpha \right) q_2^{*}$ & $\left( \gamma -\beta \right)
f $ & $\,\,\,\,\,\,\,\,\,\,\,0$%
\end{tabular}
\right) ,\,  \tag{2.12a}
\end{equation}
and 
\begin{equation}
U\left( t,x\right) \,=\,-2k_{+_t}\,=-2\left( 
\begin{tabular}{lll}
$a_t$ & $q_{1_t}$ & $q_{2_t}$ \\ 
$\pm q_{1_t}^{*}$ & $d_t$ & $e_t$ \\ 
$\pm q_{2_t}^{*}$ & $f_t$ & $g_t$%
\end{tabular}
\right) \,.  \tag{2.12b}
\end{equation}
Based on the solution of equation $\Delta ^{(2)}\left( I\,+\,\Phi
_{+}\right) \,=\,\left( I\,+\,\Phi _{+}\right) \Delta _0^{(2)}\,$,\thinspace
we get that $k_{+}\left( t,z;x\right) $ must obey the following equation: 
\[
\left( 
\begin{tabular}{lll}
$\alpha $ & $0$ & $0$ \\ 
$0$ & $\beta $ & $0$ \\ 
$0$ & $0$ & $\gamma $%
\end{tabular}
\right) k_{+t}\,+\,k_{+z}\left( 
\begin{tabular}{lll}
$\alpha $ & $0$ & $0$ \\ 
$0$ & $\beta $ & $0$ \\ 
$0$ & $0$ & $\gamma $%
\end{tabular}
\right)
\,\,\,\,\,\,\,\,\,\,\,\,\,\,\,\,\,\,\,\,\,\,\,\,\,\,\,\,\,\,\,\,\,\,\,\,\,\,%
\,\,\,\,\,\,\,\,\,\,\,\,\,\,\,\,\,\,\,\,\,\,\,\,\,\,\, 
\]
\begin{equation}
+\left( 
\begin{tabular}{lll}
$\,\,\,\,\,\,\,\,\,\,\,\,\,\,\,\,\,0$ & $\left( \alpha -\beta \right) q_1$ & 
$\left( \alpha -\gamma \right) q_2$ \\ 
$\pm \left( \beta -\alpha \right) q_1^{*}$ & $\,\,\,\,\,\,\,\,\,\,\,\,\,0$ & 
$\left( \beta -\gamma \right) e$ \\ 
$\pm \left( \gamma -\alpha \right) q_2^{*}$ & $\left( \gamma -\beta \right)
f $ & $\,\,\,\,\,\,\,\,\,\,\,0$%
\end{tabular}
\right) k_{+}=0  \tag{2.13}
\end{equation}
and if we evaluate this eq.$(2.13)\,\,$on $z\,=\,t$, we find 
\begin{eqnarray}
a_t\, &=&\,\mp \frac 1\alpha \left[ \left( \alpha -\beta \right) \mid
q_1\mid ^2+\left( \alpha -\gamma \right) \left| q_2\right| ^2\right] ,\,\,\,
\tag{2.14a} \\
d_t &=&-\frac 1\beta \left[ \left( \beta -\gamma \right) ef\pm \left( \beta
-\alpha \right) \left| q_1\right| ^2\right] ,  \tag{2.14b} \\
e_t &=&-\frac 1\beta \left[ \left( \beta -\gamma \right) eg\pm \left( \beta
-\alpha \right) q_1^{*}q_2\right] ,  \tag{2.14c} \\
f_t &=&-\frac 1\gamma \left[ \left( \gamma -\beta \right) fd\pm \left(
\gamma -\alpha \right) q_2^{*}q_1\right] ,  \tag{2.14d} \\
g_t &=&-\frac 1\gamma \left[ \left( \gamma -\beta \right) ef\pm \left(
\gamma -\alpha \right) \left| q_2\right| ^2\right] .  \tag{2.14e}
\end{eqnarray}
After substituting the above equations into eq.$(2.12b)$ and eq.$(2.2a),\,$%
we find the following equation 
\begin{equation}
\Delta ^{(1)}\,=\,I\left( i\alpha \frac \partial {\partial x}\,-\,\frac{%
\partial ^2}{\partial t^2}\right) -2\left( 
\begin{tabular}{lll}
$\,\,\,\,\,\,\,a_t$ & $q_{1_t}$ & $q_{2_t}$ \\ 
$\,\pm q_{1_t}^{*}$ & $d_t$ & $e_t$ \\ 
$\,\pm q_{2_t}^{*}$ & $f_t$ & $g_t$%
\end{tabular}
\right) .  \tag{2.15}
\end{equation}
On the other hand, by substituting eq.$(2.12a)\,\,$into\thinspace
eq.\thinspace $(2.2b)$, we get\thinspace \thinspace \thinspace $%
\,\,\,\,\,\,\,\,\,\,\,\,\,\,\,\,\,\,\,\,\,\,\,\,\,\,\,\,\,\,\,\,\,\,\,\,\,\,%
\,\,\,\,\,\,\,\,\,\,\,\,\,\,\,\,\,\,\,\,\,\,\,\,\,\,\,\,\,\,\,\,\,\,\,\,\,\,%
\,\,\,\,\,\,\,\,\,\,\,\,\,\,\,\,\,\,\,\,\,\,\,\,\,\,\,\,\,\,\,\,\,\,\,\,$ 
\[
\Delta ^{(2)}\,=\,\left( 
\begin{tabular}{lll}
$\alpha $ & $0$ & $0$ \\ 
$0$ & $\beta $ & $0$ \\ 
$0$ & $0$ & $\gamma $%
\end{tabular}
\right) \frac \partial {\partial
t}\,\,\,\,\,\,\,\,\,\,\,\,\,\,\,\,\,\,\,\,\,\,\,\,\,\,\,\,\,\,\,\,\,\,\,\,\,%
\,\,\,\,\,\,\,\,\,\,\,\,\,\,\,\,\,\,\,\,\,\,\,\,\,\,\,\,\,\,\,\,\,\,\,\,\,\,%
\,\,\,\,\,\,\,\,\,\,\,\,\,\,\,\,\,\,\,\,\,\,\,\,\,\,\,\,\,\,\,\,\,\,\,\,\,\,%
\,\,\,\,\,\,\,\,\,\,\,\,\,\,\,\,\,\,\,\,\,\,\,\,\,\,\,\,\,\,\,\,\,\,\,\,\,\,%
\,\,\,\,\,\,\,\,\,\,\,\,\,\,\,\,\,\,\,\,\,\,\,\,\,\,\,\,\,\,\,\,\,\,\,\,\,\,%
\,\,\,\,\,\,\,\,\,\,\,\,\,\,\,\,\,\,\,\,\,\,\,\,\,\,\,\,\,\,\,\,\,\,\,\,\,\,%
\,\,\,\,\,\,\,\,\,\,\,\,\,\,\,\,\,\,\,\,\,\,\,\,\,\,\,\,\,\,\,\,\,\,\, 
\]
\begin{equation}
\,\,\,\,\,\,\,\,\,\,\,\,\,\,\,\,\,\,\,\,\,\,+\left( 
\begin{tabular}{lll}
$\,\,\,\,\,\,\,\,\,\,\,\,\,\,\,\,0$ & $\left( \alpha -\beta \right) q_1$ & $%
\left( \alpha -\gamma \right) q_2$ \\ 
$\pm \left( \beta -\alpha \right) q_1^{*}$ & $\,\,\,\,\,\,\,\,\,\,\,\,\,0$ & 
$\left( \beta -\gamma \right) e$ \\ 
$\pm \left( \gamma -\alpha \right) q_2^{*}$ & $\left( \gamma -\beta \right)
f $ & $\,\,\,\,\,\,\,\,\,\,\,\,0$%
\end{tabular}
\right)
\,.\,\,\,\,\,\,\,\,\,\,\,\,\,\,\,\,\,\,\,\,\,\,\,\,\,\,\,\,\,\,\,\,\,\,\,\,%
\,\,\,\,\,\,\,\,\,\,\,\,\,\,\,\,\,\,\,\,\,\,\,\,\,\,\,\,\,\,\,\,\,\,\,\,\,\,%
\,\,\,\,\,\,\,\,\,\,\,\,\,\,\,\,\,\,\,  \tag{2.16}
\end{equation}
$\,\,\,\,\,\,\,\,$Since $\Delta ^{(1)}\,$commutes with $\Delta ^{(2)}\,,$%
\begin{eqnarray*}
iq_{1_x}\,+\,\frac{\left( \alpha +\beta \right) }{\alpha \left( \alpha
-\beta \right) }q_{1_{tt}}\,\pm 2\left( \frac 1{\alpha \left( \alpha -\beta
\right) }\right) \left( \left[ \theta _1\mid q_1\mid ^2+\theta _2\left|
q_2\right| ^2\right] q_1+Q_1\right) \, &=&\,0 \\
iq_{2_x}\,+\,\frac{\left( \alpha +\beta \right) }{\alpha \left( \alpha
-\beta \right) }q_{2_{tt}}\,\pm 2\left( \frac 1{\alpha \left( \alpha -\beta
\right) }\right) \left( \left[ \theta _3\mid q_1\mid ^2+\theta _4\left|
q_2\right| ^2\right] q_2+Q_2\right) \, &=&\,0,
\end{eqnarray*}
\begin{equation}
\tag{2.17a}
\end{equation}
and its complex conjugate 
\begin{eqnarray*}
-iq_{1_x}^{*}\,+\,\frac{\left( \alpha +\beta \right) }{\alpha \left( \alpha
-\beta \right) }q_{1_{tt}}^{*}\,\pm \frac 2{\alpha \left( \alpha -\beta
\right) }\left( \left[ \theta _5\mid q_1\mid ^2+\theta _6\left| q_2\right|
^2\right] q_1^{*}+Q_1^{*}\right) \, &=&\,0 \\
-iq_{2_x}^{*}\,+\,\frac{\left( \alpha +\beta \right) }{\alpha \left( \alpha
-\beta \right) }q_{2_{tt}}^{*}\,\pm \frac 2{\alpha \left( \alpha -\beta
\right) }\left( \left[ \theta _7\mid q_1\mid ^2+\theta _8\left| q_2\right|
^2\right] q_2^{*}+Q_2^{*}\right) \, &=&\,0,
\end{eqnarray*}
\begin{equation}
\tag{2.17b}
\end{equation}
where 
\begin{eqnarray}
\theta _1 &=&\theta _5=\frac 1{\alpha \beta }\left[ \left( \alpha +\beta
\right) \left( \alpha -\beta \right) ^2\right] ,\,\,  \tag{2.18a} \\
\theta _2 &=&\frac 1{\alpha \gamma }\left[ \gamma \left( \alpha -\beta
\right) \left( \alpha -\gamma \right) +\alpha \left( \alpha -\gamma \right)
^2\right] ,  \tag{2.18b} \\
\,\theta _3 &=&\frac 1{\alpha \beta }\left[ \beta \left( \alpha -\beta
\right) \left( \alpha -\gamma \right) +\alpha \left( \alpha -\beta \right)
^2\right] ,  \tag{2.18c} \\
\,\theta _4 &=&\theta _8=\frac 1{\alpha \gamma }\left[ \left( \alpha +\gamma
\right) \left( \alpha -\gamma \right) ^2\right] ,  \tag{2.18d} \\
\theta _6 &=&\frac 1{\alpha \beta }\left[ \left( \alpha +\beta \right)
\left( \alpha -\beta \right) \left( \alpha -\gamma \right) \right] , 
\tag{2.18e} \\
\theta _7 &=&\frac 1{\alpha \gamma }\left[ \left( \alpha +\gamma \right)
\left( \alpha -\beta \right) \left( \alpha -\gamma \right) \right] , 
\tag{2.18f} \\
Q_1 &=&\left[ \frac{\left( \alpha -\beta \right) \left( \beta -\gamma
\right) }\beta efq_1+\left( \gamma -\beta \right) fq_{2_t}-\frac{\left(
\alpha -\gamma \right) \left( \gamma -\beta \right) }\gamma fdq_2\right] , 
\nonumber \\
&&  \tag{2.18g} \\
Q_2 &=&\left[ \frac{\left( \alpha -\gamma \right) \left( \gamma -\beta
\right) }\gamma efq_2+\left( \beta -\gamma \right) eq_{1_t}-\frac{\left(
\alpha -\beta \right) \left( \beta -\gamma \right) }\beta egq_1\right] , 
\nonumber \\
&&  \tag{2.18h} \\
e &=&f^{*}\text{\thinspace and }d=\frac \gamma \beta g^{*}.  \tag{2.18i}
\end{eqnarray}
If we put parameters $\beta =\gamma =\delta $ (where $\delta $ is an
arbitrary real parameter)$\ $into the above equations and then substitute
the results into eq.$(2.17a)\,$and ($2.17b$), we find 
\begin{eqnarray}
iq_{1_x}\,+\,\frac{\left( \alpha +\delta \right) }{\alpha \left( \alpha
-\delta \right) }q_{1_{tt}}\,+2\mu \left[ \mid q_1\mid ^2+\left| q_2\right|
^2\right] q_1\, &=&\,0  \nonumber \\
i\,q_{2_x}+\frac{\left( \alpha +\delta \right) }{\alpha \left( \alpha
-\delta \right) }\,q_{2_{tt}}\,+2\mu \left[ \mid q_1\mid ^2+\left|
q_2\right| ^2\right] q_2\, &=&\,0,  \tag{2.19a}
\end{eqnarray}
and its complex conjugate 
\begin{eqnarray}
-iq_{1_x}^{*}+\frac{\left( \alpha +\delta \right) }{\alpha \left( \alpha
-\delta \right) }q_{1_{tt}}^{*}+2\mu \left[ \left| q_1\right| ^2+\left|
q_2\right| ^2\right] q_1^{*} &=&0  \nonumber \\
-iq_{2_x}^{*}+\frac{\left( \alpha +\delta \right) }{\alpha \left( \alpha
-\delta \right) }q_{2_{tt}}^{*}+2\mu \left[ \left| q_1\right| ^2+\left|
q_2\right| ^2\right] q_2^{*} &=&0.  \tag{2.19b}
\end{eqnarray}
Here parameter $\mu \,$is an arbitrary real value which has been defined in
eq.$\left( 1.2\right) $. It is obvious that eq.$(2.19a)\,$and $\left(
2.19b\right) $ are the general form of the integrable coupled nonlinear
Schr\"{o}dinger equation of Manakov type (eq.$(1.1)$) and its complex
conjugate.$\,$We note that the functions $Q_1$ and $Q_2$ in eq.($2.17a$)$\,$
and its complex conjugates in eq.$(2.17b)$ are the perturbative terms
(inhomogeneous terms) for arbitrary $\alpha $,$\,\,\beta \,\,$and $\,\gamma $%
.

\section{The Bright and Dark Multi-Soliton Solution}

We consider a general matrix function\thinspace \thinspace $F\,\,$in eq.($%
2.7b$)\thinspace and$\,$substitute\thinspace \thinspace it \thinspace into
eq.$(2.11a)$, we find four differential equations 
\begin{eqnarray}
i\alpha A_{n_x}\,\,+A_{n_{zz}}\,-\,A_{n_{tt}}\, &=&\,0,  \tag{3.1a} \\
i\alpha B_{n_x}\,\,+B_{n_{zz}}\,-\,B_{n_{tt}}\, &=&\,0,  \tag{3.1b} \\
i\alpha A_{n_x}^{*}\,\,+A_{n_{zz}}^{*}\,-\,A_{n_{tt}}^{*}\, &=&\,0, 
\tag{3.1c} \\
i\alpha B_{n_x}^{*}\,+\,B_{n_{zz}}^{*}\,-\,B_{n_{tt}}^{*}\, &=&\,0\,. 
\tag{3.1d}
\end{eqnarray}
$\,$The solution of the above$\,\,$equations$\,\,$can be derived by using
separable variable method. We then find 
\begin{eqnarray}
A_n\left( t,z;x\right) \, &=&\,\sum\limits_{n=1}^NA_{n_0}e^{-\alpha \rho
_nz}\left[ e^{\rho _n\left( \delta t+i\rho _n\left( \alpha ^2-\delta
^2\right) x\right) }\right] ,  \tag{3.2a} \\
B_n\left( t,z;x\right)  &=&\sum\limits_{n=1}^NB_{n_0}e^{-\alpha \rho
_nz}\left[ e^{\rho _n\left( \delta t+i\rho _n\left( \alpha ^2-\delta
^2\right) x\right) }\right] ,  \tag{3.2b} \\
A_n^{*}\left( t,z;x\right)  &=&\sum\limits_{n=1}^NA_{n_0}^{*}e^{-\delta
\sigma _nz}\left[ e^{\sigma _n\left( \alpha t+i\sigma _n\left( \delta
^2-\alpha ^2\right) x\right) }\right] ,  \tag{3.2c} \\
B_n^{*}\left( t,z;x\right) \, &=&\,\sum\limits_{n=1}^NB_{n_0}^{*}e^{-\delta
\sigma _nz}\left[ e^{\sigma _n\left( \alpha t+i\sigma _n\left( \delta
^2-\alpha ^2\right) x\right) }\right] ,  \tag{3.2d}
\end{eqnarray}
where $A_{n_0}\,$\thinspace , $A_{n_0}^{*}$, $B_{n_0}\,$\thinspace
and\thinspace $B_{n_0}^{*}\,\,$are arbitrary complex parameters.\thinspace
\thinspace $\,\,\,\,\,\,\,\,\,\,\,\,\,\,\,\,\,\,\,\,\,\,\,\,\,\,\,\,\,\,\,\,%
\,\,\,\,\,\,\,\,\,\,\,\,\,\,\,\,\,\,\,\,\,\,\,\,\,\,\,\,\,\,\,\,\,\,\,\,\,\,%
\,\,\,\,\,\,\,\,\,\,\,\,\,\,\,\,\,\,\,\,\,\,\,\,\,\,\,\,\,\,\,\,\,\,\,\,\,\,%
\,\,\,\,\,\,\,\,\,\,\,\,\,\,\,\,\,\,\,\,\,\,\,\,$

To get the final solution of the integrable coupled \textbf{NLS} equation of
Manakov type, we have to substitute the equations $(2.7b)$, $(3.2a)$, $%
(3.2b) $, ($3.2c)\,\,$and$\,\,(3.2d)\,\,$into Marchenko matrix equation (eq.$%
(2.9a)$). We get (for $a$, $q_{1\text{,}}\,$and $q_2$) 
\[
\]
\[
a\,=\,-\sum\limits_{n=1}^N\int\limits_t^\infty q_1\,A_{n_0}^{*}e^{i\frac{%
\sigma _n^2}\alpha \left( \delta ^2-\alpha ^2\right) x}e^{-\delta \sigma
_nz}e^{\alpha \sigma _nt^{\prime }}dt^{\prime
}\,\,\,\,\,\,\,\,\,\,\,\,\,\,\,\,\,\,\,\,\,\,\,\,\,\,\,\,\,\,\,\,\,\,\,\,\,%
\,\,\,\,\,\,\, 
\]
\begin{equation}
-\sum\limits_{n=1}^N\int\limits_t^\infty q_2B_{n_0}^{*}e^{i\frac{\sigma _n^2}%
\alpha \left( \delta ^2-\alpha ^2\right) x}e^{-\delta \sigma _nz}e^{\alpha
\sigma _nt^{\prime }}dt^{\prime },  \tag{3.3a}
\end{equation}
\begin{eqnarray}
q_1\, &=&\,-\sum\limits_{n=1}^N\left( e^{-\alpha \rho _nz}A_{n_0}\right)
e^{\delta \rho _nt}e^{i\frac{\rho _n^2}\alpha \left( \alpha ^2-\delta
^2\right) x}  \nonumber \\
&&\,-\sum\limits_{n=1}^N\left( A_{n_0}\right) \int\limits_t^\infty a\left(
t,z;x\right) \,e^{i\frac{\rho _n^2}\alpha \left( \alpha ^2-\delta ^2\right)
x\,}e^{-\alpha \rho _nz}e^{\delta \rho _nt^{\prime }}dt^{\prime }\,, 
\tag{3.3b}
\end{eqnarray}
and 
\begin{eqnarray}
q_2 &=&-\sum\limits_{n=1}^N\left( e^{-\alpha \rho _nz}B_{n_0}\right)
e^{\delta \rho _nt}e^{i\frac{\rho _n^2}\alpha \left( \alpha ^2-\delta
^2\right) x}  \nonumber \\
&&-\sum\limits_{n=1}^N\left( B_{n_0}\right) \int\limits_t^\infty a\left(
t,z;x\right) e^{i\frac{\rho _n^2}\alpha \left( \alpha ^2-\delta ^2\right)
x}e^{-\alpha \rho _nz}e^{\delta \rho _nt^{\prime }}dt^{\prime }.  \tag{3.3c}
\end{eqnarray}
The final solution is work on $z=t$. Hence, by substituting eq.$(3.3a)$ to
eq.$(3.3b)\,$and eq.$(3.3c)$ we find the solution (for $n=1,2,...,N$): 
\begin{equation}
q_1\,=\frac{-A_{n_0}\,e^{\rho _n\left( \delta -\alpha \right) t}e^{-i\frac{%
\rho _n^2}\alpha \left( \delta ^2-\alpha ^2\right) x}}{1\,+\,\left( \frac{%
\mu \left( \left| A_{n_0}\right| ^2+\left| B_{n_0}\right| ^2\right) }{\left(
\left( \frac{\alpha ^2-\delta ^2}\alpha \right) ^{1/2}\left( \rho _n-\sigma
_n\right) \right) ^2}\right) e^{\rho _n\left( \delta -\alpha \right) t-i%
\frac{\rho _n^2}\alpha \left( \delta ^2-\alpha ^2\right) x}e^{-\sigma
_n\left( \delta -\alpha \right) t+i\frac{\sigma _n^2}\alpha \left( \delta
^2-\alpha ^2\right) x}}\,\,,  \tag{3.4a}
\end{equation}
and 
\begin{equation}
q_2=\frac{-B_{n_0}e^{\rho _n\left( \delta -\alpha \right) t}e^{-i\frac{\rho
_n^2}\alpha \left( \delta ^2-\alpha ^2\right) x}}{1+\left( \frac{\mu \left(
\left| A_{n_0}\right| ^2+\left| B_{n_0}\right| ^2\right) }{\left( \left( 
\frac{\alpha ^2-\delta ^2}\alpha \right) ^{1/2}\left( \rho _n-\sigma
_n\right) \right) ^2}\right) e^{\rho _n\left( \delta -\alpha \right) t-i%
\frac{\rho _n^2}\alpha \left( \delta ^2-\alpha ^2\right) x}e^{-\sigma
_n\left( \delta -\alpha \right) t+i\frac{\sigma _n^2}\alpha \left( \delta
^2-\alpha ^2\right) x}}.  \tag{3.4b}
\end{equation}
We define

\begin{equation}
\eta _n\,=\,k_n\left( t+ik_n\xi x\right) ,  \tag{3.5a}
\end{equation}
and 
\begin{equation}
\eta _n^{*}\,=\,k_n^{*}\left( t-ik_n^{*}\xi x\right) \,\,\,,\,  \tag{3.5b}
\end{equation}
where 
\begin{equation}
k_n\,=\,\left( \delta -\alpha \right) \rho _n\,\,,\,  \tag{3.5c}
\end{equation}
\begin{equation}
k_n^{*}\,=\,-\left( \delta -\alpha \right) \sigma _n\,\,,  \tag{3.5d}
\end{equation}
\begin{equation}
\tau =\left( \frac{\alpha ^2-\delta ^2}\alpha \right) ^{1/2}\left( \frac
1{\delta -\alpha }\right) ,  \tag{3.5e}
\end{equation}
and

\begin{equation}
\xi =\frac{\left( \alpha ^2-\delta ^2\right) }{\alpha \left( \delta -\alpha
\right) ^2}.  \tag{3.5f}
\end{equation}
Here $\,\rho _n\,\,$and$\,\,\sigma _n\,\,$are\thinspace arbitrary\thinspace
complex\thinspace parameters\thinspace and$\,\,\rho _n^{*}\,=-\sigma _n.$

Now $q_1\,$and $q_2$\thinspace \thinspace can be rewritten as 
\begin{equation}
q_1\,=\sum\limits_{n=1}^N\frac{-A_{n_0}\,e^{\eta _n}}{1\,+\,e^{R_n\,+\,\eta
_n\,+\,\eta _n^{*}}}\,,  \tag{3.6a}
\end{equation}
and 
\begin{equation}
q_2\,=\sum\limits_{n=1}^N\frac{-B_{n_0}\,e^{\eta _n}}{1\,+\,e^{R_n\,+\,\eta
_n\,+\,\eta _n^{*}}},  \tag{3.6b}
\end{equation}
where 
\begin{equation}
e^{R_n}\,=\frac{\mu \left( \left| A_{n_0}\right| ^2+\left| B_{n_0}\right|
^2\right) }{\left( \tau k_n\,+\,\tau k_n^{*}\right) ^2}.  \tag{3.7}
\end{equation}
Based on our solution in eq.$(3.6a)$ and $(3.6b)$, we can see that our
results are the bright and dark multi-soliton solution since $\mu =\pm \frac{%
\alpha ^2-\delta ^2}{\alpha ^2\delta }$.The results of $q_1$and $q_2$ show
that there are an collision of the bright and dark multi-soliton. If we
choose parameter $\left( \left( \frac{\left( \alpha +\delta \right) }{\alpha
\left( \alpha -\delta \right) }\right) \mu \right) $\thinspace $>0$ then we
get the bright $N$-soliton solution. On the other hand, the dark $N$-soliton
solution is found when $\left( \left( \frac{\left( \alpha +\delta \right) }{%
\alpha \left( \alpha -\delta \right) }\right) \mu \,\right) <\,0$\thinspace
. In the equations, there are only four arbitrary complex
parameters\thinspace $A_{n_0}\,$, $B_{n_0}$, $\rho _n\,\,$and$\,\,\sigma _n$ 
$\,$which can directly influence the phase of the solitons interaction. The
results of $q_1$ and $q_2$ can also be rewritten in the more conventional
form by introducing $\rho _n\,=l_n+i\lambda _n$ (where $l_n$ and $\lambda _n$
are arbitrary real parameters), 
\begin{equation}
q_1\,=\sum\limits_{n=1}^N\frac{\left( \frac{-\left( \frac{\alpha ^2-\delta ^2%
}\alpha \right) ^{1/2}l_nA_{n_0}}{\sqrt{\mu \left( \left| A_{n_0}\right|
^2+\left| B_{n_0}\right| ^2\right) }}\right) \exp \left( i\left( \delta
-\alpha \right) \left[ \lambda _nt\,+\,\xi \left( \delta -\alpha \right)
\left( l_n^2-\lambda _n^2\right) x\right] \right) }{\cosh \left[ \left(
\delta -\alpha \right) l_n\left( t-2\xi \left( \delta -\alpha \right)
\lambda _nx\,\right) +\,\varphi _n\right] },  \tag{3.8}
\end{equation}
and 
\begin{equation}
q_2\,=\sum\limits_{n=1}^N\frac{\left( \frac{-\left( \frac{\alpha ^2-\delta ^2%
}\alpha \right) ^{1/2}l_nB_{n_0}}{\sqrt{\mu \left( \left| A_{n_0}\right|
^2+\left| B_{n_0}\right| ^2\right) }}\right) \exp \left( i\left( \delta
-\alpha \right) \left[ \lambda _nt\,+\,\xi \left( \delta -\alpha \right)
\left( l_n^2-\lambda _n^2\right) x\right] \right) \,}{\cosh \left[ \left(
\delta -\alpha \right) l_n\left( t\,-\,2\xi \left( \delta -\alpha \right)
\lambda _nx\,\right) +\,\varphi _n\right] }\,\,,  \tag{3.9}
\end{equation}
where $\varphi _n$ is a real multi-soliton phase, 
\begin{equation}
\varphi _n\,=\,\frac 12R_{n,}\,  \tag{3.10}
\end{equation}
and the amplitudes of the multi-soliton are

\begin{equation}
Amp^1=-\frac{\left( \frac{\alpha ^2-\delta ^2}\alpha \right) ^{1/2}l_nA_{n_0}%
}{\sqrt{\mu \left( \left| A_{n_0}\right| ^2+\left| B_{n_0}\right| ^2\right) }%
},  \tag{3.11a}
\end{equation}
and

\begin{equation}
Amp^2=-\frac{\left( \frac{\alpha ^2-\delta ^2}\alpha \right) ^{1/2}l_nB_{n_0}%
}{\sqrt{\mu \left( \left| A_{n_0}\right| ^2+\left| B_{n_0}\right| ^2\right) }%
}.  \tag{3.11b}
\end{equation}
Parameter $\lambda _n$ contributes to the velocities of the multi-soliton.$%
^{15}$

The results derived above have contributed on the obtaining a general form
of an exact real parameter $\mu \,$and have also shown that there is a
general multi-soliton solution of the integrable coupled \textbf{NLS} eq. of
Manakov type (eq.$(1.1)$).

\thinspace \thinspace Based on our results in eq.$(3.6a)$ and $(3.6b)$, the
bright $N$-solitons solution (for $\left( \left( \frac{\left( \alpha +\delta
\right) }{\alpha \left( \alpha -\delta \right) }\right) \mu \right) $ $>0$)
can be reduced to the bright one and two soliton solutions related
to\thinspace that in the works that have been done before by Radhakrishnan,
\thinspace et. \thinspace al.\thinspace \thinspace using Hirota method in
Ref.$10$. According to the comparison of both methods, their \thinspace
results\thinspace of\thinspace the bright one soliton solution\thinspace is
equal to our results\thinspace when$\,\alpha \,e^{\eta _1^{\left( 0\right)
}}=\,-A_{1_0}\,$,$\,\,\beta e^{\eta _1^{\left( 0\right) }}=-B_{1_0}\,$, $%
\frac{\left( \alpha +\delta \right) }{\alpha \left( \alpha -\delta \right) }%
=1$, $\xi =1$, $\tau =1$, $k_1=\left( \delta -\alpha \right) \left(
l_1+i\lambda _1\right) $,$\,\mu =+\frac{\alpha ^2-\delta ^2}{\alpha ^2\delta 
}$ and$\,\left( \mid \alpha \mid ^2+\mid \beta \mid ^2\right) =\left( \left|
A_{1_0}\right| ^2+\left| B_{1_0}\right| ^2\right) \,$.\thinspace On the
other hand,\thinspace their results of the inelastic\thinspace collision of
the bright two soliton\thinspace \thinspace can be reduced to our\thinspace
\thinspace elastic\thinspace \thinspace collision\thinspace \thinspace of
the solution if\thinspace we put $\alpha _1:\alpha _2=\beta _1:\beta _2$%
\thinspace in\thinspace their\thinspace \thinspace result.

We can also compare our results with the results in Ref.$8$. We find that
our result and the Shchesnovich result are the same when $\chi =\frac
12,\,2\mu =1,\,x=z,\,t=\tau ,$\thinspace $\,n=1\,$and $\left| \theta
_1\right| ^2+\left| \theta _2\right| ^2=\left| A_{1_0}\right| ^2+\left|
B_{1_0}\right| ^2=1.\,$So, we can generalize that our result is the solution
of the bright and dark multi-soliton collisions.

\section{Discussions and Conclusions}

We have presented a general form of the bright and dark multi-soliton
solution of the integrable coupled NLS equation\thinspace of Manakov type
using the inverse scattering Zakharov-Shabat scheme.We can conclude that the
solution of the Manakov type can be solved by using an expanded inverse
scattering Zakharov-Shabat scheme in which we have chosen a certain operator
in eq.$(2.1)$, and a certain $k_{+}\,\,$(eq.$\left( 2.7a\right) $)$\,$and $%
F\,$(eq.$\left( 2.7b\right) $) in our solution.

Finally, we find the novel result provided in Ref.$9$ and the results of the
use of Hirota method in Ref.$10$ and Rev.$11\,$that the solution in eq.$%
(3.6a)$ and $(3.6b)\,\,$corresponds to an elastic collision of the bright
and dark multi-soliton, as long as $A_{1_0}:A_{2_0}:...:A_{N_0}$ is equal to 
$B_{1_0}:B_{2_0}:...:B_{N_0}$. If in our results, $%
A_{1_0}:A_{2_0}:...:A_{N_0}\neq B_{1_0}:B_{2_0}:...:B_{N_0}$, then we get an
inelastic collision of the bright and dark multi-solitons. From the eq.$%
(3.8)\,$and$\,$eq.$(3.9)$, we can also\thinspace conclude that although the
collision between the bright and dark multi-solitons their\thinspace
velocities\thinspace and amplitudes or intensities do not change, their
phases do change.

According to the comparison between our bright one and two soliton solution
of the integrable coupled \textbf{NLS} equation of Manakov type and the
solution provided by Radhakrishnan,\thinspace R., et.al., we get that their
inelastic\thinspace collision of the\thinspace bright\thinspace two-soliton
solution can be our \thinspace elastic collision solution, \thinspace if
some\thinspace certain parameters in the results are omitted by putting $%
\alpha _1:\alpha _2=\beta _1:\beta _2$.\thinspace In our results,\thinspace
the parameter exp$\left( \eta _j^{(0)}\right) $appeared in Ref.$10$ has been
absorbed \thinspace into $A_{n_0}$ and $B_{n_0}$. Based on our results, we
propose that there must be an \thinspace integrable \thinspace $m$-tupled
(vector) multi-solitons\thinspace solution of the following equation:$%
^{4,16,17}$ 
\[
\left( i\frac \partial {\partial x}+\chi \frac{\partial ^2}{\partial t^2}%
+2\mu \sum\limits_{b=1}^m\left| q_b\right| ^2\right)
q_c\,=\,0,\,\,\,c=1,2,...,m,
\]
where $\chi \,$and$\,\mu $ are arbitrary real parameters. We also propose
that there must be an elastic and inelastic collision of the bright and dark
multi-soliton in this system.$^{17}$

$\mathbf{ACKNOWLEDGEMENTS}$

Both authors would like to thank H.J.Wospakrik for useful discussions. We
also thank Yu. S. Kivshar and N.N. Akhmediev (Optical Sciences Centre, The
Institute for Advanced Studies, The Australian National University) $\,$for
their information and suggestions. We also would like to thank P. Silaban
\thinspace for his encouragements. The work of F.Z. is supported by the
Hibah Bersaing VII/1, 1998-1999, DIKTI, Republic of Indonesia.The work of
H.E. is partially supported by CIDA - EIUDP, Republic of Indonesia.

\end{document}